\documentclass[%
preprint,
showkeys,
amsmath,amssymb,
aps,
]{revtex4-1}

\usepackage{graphicx}
\usepackage{dcolumn}
\usepackage{bm}

\begin{document}

\preprint{Journal of Gravity}

\title{Collapse of a relativistic self-gravitating star with radial heat flux: Impact of anisotropic stresses} 

\author{Ranjan Sharma}
\altaffiliation[Previously at: ]{School of Physics (Pietermaritzburg Campus), University of KwaZulu-Natal, South Africa;\\
Department of Physics, St. Joseph's College, Darjeeling, West Bengal, India.}
\email{rsharma@iucaa.ernet.in}
\author{Shyam Das}%
\email{dasshyam321@gmail.com}
\affiliation{%
P. D. Women's College, Jalpaiguri 735101, West Bengal, India.}


\date{\today}

\begin{abstract}
We develop a simple model for a self-gravitating spherically symmetric relativistic star which begins to collapse from an initially static configuration by dissipating energy in the form of radial heat flow. We utilize the model to show how local anisotropy effects the collapse rate and thermal behaviour of gravitationally evolving systems.
\end{abstract}

\maketitle


\section{\label{sec1} Introduction}
In cosmology and astrophysics, there exist many outstanding issues relating to a dynamical system collapsing under the influence of its own gravity. In view of Cosmic Censorship Conjecture, the general relativistic prediction is that such a collapse must terminate into a space-time singularity covered under its event horizon though there are several counter examples where it has been shown that a naked singularity is more likely to be formed (see \cite{Joshi01} and references therein). In astrophysics, the end stage of a massive collapsing star has long been very much speculative in nature\cite{Chandra,Joshi01}. From classical gravity perspective, to get a proper understanding of the nature of collapse and physical behaviour of a collapsing system, construction of a realistic model of the collapsing system is necessary. This, however, turns out to be a difficult task because of the highly non-linear nature of the governing field equations. To reduce the complexity, various simplifying methods are often adopted and the pioneering work of Oppenheimer and Snyder\cite{OppenS} was a first step in this direction when collapse of a highly idealized spherically symmetric dust cloud was studied. Since then, various attempts have been made to develop realistic models of gravitationally collapsing systems to understand the nature and properties of collapsing objects. It got a tremendous impetus when Vaidya\cite{Vaidya} presented a solution describing the exterior gravitational field of a stellar body with outgoing radiation and Santos\cite{Santos} formulated the junction conditions joining the interior space-time of the collapsing object to the Vaidya exterior metric\cite{Vaidya}. These developments have enabled many investigators to construct realistic models of gravitationally evolving systems and also to analyze critically relevance of various factors such as shear, density inhomogeneity, local anisotropy, electromagnetic field, viscosity etc., on the physical behaviour of collapsing bodies\cite{Bonnor,Oliveira1,Oliveira2,Herrera1,Herrera2,Herrera3,Herrera4,Herrera5,Herrera6,Herrera7,Herrera8,Herrera9,Prisco1,Prisco2,Chan1,Chan2,Chan3,Chan4,Tikekar1,Maharaj1,Sharif1,Sharif2,Sharif3,Sharif4,Sharif5,Barreto1,Ghezzi1,Goswami1,Chak1,Eard1,Mena1,Thiru1,Govinder1,Dirk1,Iva1,Sarwe1,Sharma1,Sharma2}. In the absence of any established theory governing gravitational collapse, such investigations have been found to be very useful to get a proper understanding about systems undergoing gravitational collapse. 

The aim of the present work is to develop a simple model of a collapsing star and investigate the impact of pressure anisotropy on the overall behaviour of the collapsing body. Anisotropic stresses may occur in astrophysical objects for various reasons which include phase transition, density inhomogeneity, shear, electromagnetic field etc.\cite{Bowers,Herrera1,Iva2}. In \cite{Iva2}, it has been shown that influences of shear, electromagnetic field etc. on self-bound systems can be absorbed if the system is considered to be anisotropic, in general. Local anisotropy has been a well motivated factor in the studies of astrophysical objects and its role on the gross features of static stellar configurations have been investigated by many authors (see for example, Ref.~\cite{Herrera1,Herrera2,Iva3,Mak,Harko,Dev,Karma} and references therein). For dynamical systems, though pressure anisotropy is, in general, incorporated in the construction, very few works have been reported till date where impacts of anisotropy have been discussed explicitly\cite{Herrera3,Herrera4,Herrera5,Martinez,Chan3}. Appropriate junction conditions for an anisotropic fluid collapsing on the background space-time described by the Vaidya metric has been obtained in \cite{Chan4}. Considering a spheroidal geometry of Vaidya and Tikekar\cite{Vaidya2}, Sarwe and Tikekar\cite{Sarwe1} have analyzed the impact of geometry vis-a-vis matter composition on the collapse of stellar bodies which begin to collapse from initial static configurations possessing equal compactness. Sharma and Tikekar\cite{Sharma1,Sharma2} have investigated the evolution of non-adiabatic collapse of a shear-free spherically symmetric object with anisotropic stresses on the background
of space-time obtained by introducing an inhomogeneous perturbation in the Robertson-Walker space–time.

In the present work, we have developed a model describing a shear-free spherically symmetric fluid distribution radiating away its energy in the form of radial heat flux. The star begins its collapse from an initially static configuration whose energy-momentum tensor describing the material composition has been assumed to be anisotropic, in general. To develop the model of the initial static star, we have utilized the Finch and Skea\cite{Finch} ansatz which has earlier been found to be useful to develop physically acceptable models capable of describing realistic stars in equilibrium\cite{Hans,Tik01,Tik02,Jotania}. The back ground space-time for a static configuration for the given ansatz has a clear geometrical interpretation as may be found in Ref.~\cite{Tik02}. By assuming a particular form of the anisotropic parameter, we have solved the relevant field equations and constructed a model for the initial static stellar configuration which could either be isotropic or anisotropic in nature. The solution provided by Finch and Skea\cite{Finch} is a sub-class of the solution provided here. Since, the solution presented here provides a wider range of values of the anisotropic parameter, it enables us to examine the impact of anisotropic stresses on the evolution of a large class of initial static configurations. 

Our paper has been organized as follows: In Section~\ref{sec2}, we have presented the basic equations governing the system undergoing non-adiabatic radiative collapse. In Section~\ref{sec3}, by assuming a particular anisotropic profile, we have solved the relevant field equations to develop a model for the initial static star. In Section~\ref{sec4}, by stipulating the boundary conditions across the surface separating the stellar configuration from the Vaidya\cite{Vaidya} space-time, we have solved the surface equation which governs the evolution of the initial static star that begins to collapse when the equilibrium is lost. In Section~\ref{sec5}, we have analyzed the impact of anisotropy on gravitationally collapsing systems by considering evolution of initial static stars which could either be isotropic or anisotropic. Impact of anisotropy on the evolution of temperature has been analyzed in Section~\ref{sec6}. Finally, some concluding remarks have been made in Section~\ref{sec7}.

\section{\label{sec2} Equations governing the collapsing system}
We write the line element describing the interior space-time of a spherically symmetric star collapsing under the influence of self-gravity (in standard coordinates $x^0 = t$,  $x^1=r$,  $x^2 = \theta$,  $x^3 = \phi$) as
\begin{equation}
ds_{-}^2 = -A_{0}^2(r)dt^2 + f^2(t)[B_{0}^2(r)dr^2 + r^2(d\theta^2 + \sin^2\theta d\phi^2)],\label{eq1}
\end{equation}
where, $A_{0}(r)$, $B_{0}(r)$ and $f(t)$ are yet to be determined. Note that in Eq.~(\ref{eq1}), if we set $f(t) = 1$, the metric corresponds to a static spherically symmetric configuration. 

We assume that the matter distribution of the collapsing object is an imperfect fluid described by an energy-momentum tensor of the form 
\begin{equation}
T_{\alpha\beta} = (\rho + p_t)u_{\alpha} {u_\beta} + p_{t} g_{\alpha \beta} + (p_r - p_t)\chi_{\alpha} \chi_{\beta} + q_{\alpha} u_{\beta} + q_{\beta}u_{\alpha},\label{eq2}
\end{equation}
where, $\rho$ represents the energy-density, $p_r$ and $p_t$, respectively denote fluid pressures along the radial and transverse directions, $u^\alpha$ is the $4$-velocity of the fluid, $\chi^\alpha$ is a unit space-like $4$-vector along the radial direction and $q^\alpha = (0,q^1,0,0)$ is the heat flux vector which is orthogonal to the velocity vector so that $u^\alpha u_\alpha = -1$ and $u^\alpha q_\alpha = 0$.

The Einstein's field equations governing the evolution of the system are then obtained as (we set  $G = c = 1$)
\begin{eqnarray}
8\pi\rho &=& \frac{1}{f^2}\left[\frac{1}{r^2}-\frac{1}{r^2 B_0^2}+\frac{2B_0'}{r B_0^3}\right] + \frac{3\dot{f}^2}{A_0^2 f^2},\label{eq3}\\
8\pi p_r &=& \frac{1}{f^2}\left[-\frac{1}{r^2}+\frac{1}{B_0^2 r^2}+\frac{2A_0'}{r A_0B_0^2}\right] - \frac{1}{A_0^2}\left[2\frac{\ddot{f}}{f}+\frac{\dot{f}^2}{f^2}\right],\label{eq4}\\
8\pi p_t &=& \frac{1}{f^2}\left[\frac{A_0''}{A_0B_0^2} + \frac{A_0'}{rA_0B_0^2} - \frac{B_0'}{r B_0^3} - \frac{A_0'B_0'}{A_0 B_0^3}\right] - \frac{1}{A_0^2}\left[2\frac{\ddot{f}}{f} + \frac{\dot{f}^2}{f^2}\right],\label{eq5}\\
8\pi q^1 &=& -\frac{2A_0'\dot{f}}{A_0^2 B_0^2f^3}.\label{eq6}
\end{eqnarray}
In Eqs.~(\ref{eq3})-(\ref{eq6}), a `prime' denotes differentiation with respect to $r$ and a `dot' denotes differentiation with respect to $t$. Making use of Eqs.~(\ref{eq4}) and (\ref{eq5}), we define the anisotropic parameter of the collapsing object as
\begin{equation}
\Delta(r,t) = 8\pi (p_r-p_t) = \frac{1}{f^2}\left[-\frac{A_0''}{A_0B_0^2} + \frac{A_0'}{r A_0B_0^2} + \frac{B_0'}{r B_0^3} + \frac{A_0'B_0'}{A_0B_0^3} + \frac{1}{r^2B_0^2} - \frac{1}{r^2}\right].\label{eq7} 
\end{equation}
We rewrite Eqs.~(\ref{eq3})-(\ref{eq5}) as
\begin{eqnarray}
8\pi\rho &=&\frac{8\pi\rho_s }{f^2} + \frac{3\dot{f}^2}{A_0^2 f^2},\label{eq8}\\
8\pi p_r &=& \frac{8\pi (p_r)_s}{f^2} - \frac{1}{A_0^2}\left[2\frac{\ddot{f}}{f}+\frac{\dot{f}^2}{f^2}\right],\label{eq9}\\
8\pi p_t &=& \frac{8\pi (p_t)_s}{f^2} - \frac{1}{A_0^2}\left[2\frac{\ddot{f}}{f}+\frac{\dot{f}^2}{f^2}\right],\label{eq10} 
\end{eqnarray}
where $\rho_s$, $(p_r)_s$ and $(p_t)_s$ denote the energy-density, radial pressure and tangential pressure, respectively of the initial static star. 
 
\section{\label{sec3} Interior space-time of the initially static configuration}
In our construction, we assume that an initially static star (with $f(t) = 1$ in Eq.~(\ref{eq1})), described by metric potentials $A_0(r)$, $B_0(r)$ and anisotropy $\Delta_s(r)$, starts collapsing if, for some reasons, it loses its equilibrium. To develop a model of the initially static configuration, we first assume that the anisotropic parameter is separable in its variables so that $\Delta(r,t) = \Delta_s(r)/f^2(t)$. Eq.~(\ref{eq7}) then reduces to
\begin{equation}
\Delta_s(r) = \left[-\frac{A_0''}{A_0 B_0^2} + \frac{A_0'}{r A_0B_0^2} + \frac{B_0'}{r B_0^3} +\frac{A_0'B_0'}{A_0 B_0^3}+\frac{1}{r^2B_0^2} - \frac{1}{r^2}\right],\label{eq11}
\end{equation}
which is independent of $t$. Eq.~(\ref{eq11}) can only be solved if any two of the unknown functions ($A_0$, $B_0$ and $\Delta_s(r)$) are specified. To develop a physically reasonable model of the initial static configuration, we first utilize the Finch and Skea\cite{Finch} ansatz for the metric potential $B_0$ given by
\begin{equation}
B_0^2 (r) = \left(1 +\frac{r^2}{R^2}\right),\label{eq12}
\end{equation}
where $R$ is the curvature parameter describing the geometry of the configuration. In the static case, the ansatz (\ref{eq12}) has a clear geometric characterization and has been found to yield realistic models for compact stellar objects\cite{Tik01,Tik02,Jotania}. Substituting Eq.~(\ref{eq12}) in Eq.~(\ref{eq11}), we obtain,
\begin{equation}
\left(\frac{1-x^2}{R^2 x^4}\right)\frac{d^2A_0}{dx^2} - 2\left(\frac{1 - x^2}{R^2 x^5}\right)\frac{dA_0}{dx} + \left[\left(\frac{1 - x^2}{R^2 x^4}\right) - \Delta_s\right]A_0 = 0,\label{eq13}
\end{equation} 
where we have used the following transformation:
\begin{equation}
x^2 = \left(1 + \frac{r^2}{R^2}\right).\label{eq14}
\end{equation}
To solve Eq.~(\ref{eq13}), we assume a particular anisotropic profile 
\begin{equation}
\Delta_s (x) = \frac{\alpha(x^2-1)(2-x^2)}{R^2x^6},\label{eq15}
\end{equation}
where $\alpha$ is the measure of anisotropy. The motivation for choosing the particular form of the anisotropy parameter are the following: (1) It is physically reasonable as the anisotropy vanishes at the centre ($r=0$, i.e., $x=1$) as expected and (2) it provides a solution of Eq.~(\ref{eq13}) in closed form. Note that $\alpha = 0$ corresponds to an initial static star which is isotropic in nature. Substituting Eq.~(\ref{eq15}) in Eq.~(\ref{eq13}), we get
\begin{equation}
\frac{d^2A_0}{dx^2}-\frac{2}{x}\frac{dA_0}{dx}+\left[1-\frac{\alpha(x^2-2)}{x^2}\right]A_0 = 0,\label{eq16}
\end{equation} 
whose solution is obtained as
\begin{equation}
A_{0}(x) = P x^{3/2}J_{\frac{1}{2}\sqrt{9-8\alpha}}(-ix\sqrt{-1+\alpha}) +  Q x^{3/2}Y_{\frac{1}{2}\sqrt{9-8\alpha}}(-ix\sqrt{-1+\alpha}),\label{eq17} 
\end{equation}
where $P$ and $Q$ are integration constants; $J_{\frac{1}{2}\sqrt{9-8\alpha}}(-ix\sqrt{-1+\alpha})$ is the Bessel function of first kind of order ${\frac{1}{2}\sqrt{9-8\alpha}}$ and $Y_{\frac{1}{2}\sqrt{9-8\alpha}}(-ix\sqrt{-1+\alpha})$ is the Bessel function of second kind of order ${\frac{1}{2}\sqrt{9-8\alpha}}$. It is obvious that solution is valid for $ \alpha < 1$. At $\alpha =1$, the Bessel function encounters a singularity and, therefore, we shall deal with the $\alpha = 1$ case separately.   
\subsection*{Special cases:}
\begin{itemize}
\item Case I :  $\alpha = 0$ ($\Delta_s (r) = 0$), i.e., initial static configuration is isotropic in nature.

Eq.~(\ref{eq13}) reduces to
\begin{equation}
\frac{d^2A_0}{dx^2}-\frac{2}{x}\frac{dA_0}{dx} + A_0 = 0,\label{eq18}
\end{equation} 
whose solution is found to be
\begin{equation}
A_0(x) = [(C-D x)\cos x + (C x+D)\sin x],\label{eq19}
\end{equation}
where, $C$ and $D$ are integration constants. This particular solution has been found previously in Ref.~\cite{Finch}. However, the solution (\ref{eq19}) can be obtained directly from the solution (\ref{eq17}) by substituting $\alpha = 0$ and setting $C= -\sqrt{\frac{2}{\pi}}Q$, and $D=\sqrt{\frac{2}{\pi}}P$.   

\item Case II :  $\alpha = 1$ ($\Delta_s (r) \neq 0$).

Eq.~(\ref{eq16}) reduces to 
\begin{equation}
\frac{d^2A_0}{dx^2}-\frac{2}{x}\frac{dA_0}{dx}+\frac{2}{x^2}A_0 = 0,\label{eq20}
\end{equation} 
whose solution is given by 
\begin{equation}
A_0(x)= A x + B x^2,\label{eq21}
\end{equation}
where $A$ and $B$ are integration constants\cite{Tik01}. Unfortunately, the above solution can not be regained from the general solution (\ref{eq17}) due to the  properties of Bessel functions and should be treated separately. 
\end{itemize}

We thus have a model for an initially static stellar configuration which could either be isotropic or anisotropic in nature. After loss of equilibrium, the initially static star starts collapsing and to generate a solution of the subsequent evolving system, we need to determine $f(t)$. This will be taken up in the following section.

\section{\label{sec4} Exterior space-time and junction conditions}
In our construction, evolution of the collapsing object is  governed by the function $f(t)$ which can be determined from the boundary conditions across the boundary surface joining the interior space-time and the exterior space-time described by the Vaidya\cite{Vaidya} metric 
\begin{equation}
ds_{+}^2 = -\left(1-\frac{2m(v)}{\bar{r}}\right)dv^2 - 2dvd\bar{r} + \bar{r}^2d (d\theta^2 + \sin^2\theta d\phi^2).\label{eq22}
\end{equation}
In Eq.~(\ref{eq22}), $v$ denotes the retarded time and $m(v)$ represents the total mass of the collapsing star. The junction conditions are determined by assuming a time-like $3$-surface $\Sigma$ which separates the interior and the exterior manifolds\cite{Santos}. Continuity of the metric space-times ($(ds_{-}^2)_{\Sigma} = (ds_{+}^2)_{\Sigma} = ds_{\Sigma}^2$) and the extrinsic curvatures ($K_{ij}^{-} = K_{ij}^{+}$) across the surface $\Sigma$, then yield the following matching conditions linking smoothly the interior ($r \leq r_\Sigma$) and the exterior ($ r \geq r_\Sigma$) space-times across the boundary:
\begin{eqnarray}
A_0(r_{\Sigma})dt &=& \left(1-\frac{2m}{\bar{r}}+2\frac{d{\bar{r}}}{dv}\right)^{\frac{1}{2}}_{\Sigma} dv, \label{eq23}\\
r_{\Sigma}f(t) &=& \bar{r}_{\Sigma}(v), \label{eq24}\\
m(v) &=& \left(\frac{rf(t)}{2}\left[\frac{r^2}{R^2+r^2}+\left(\frac{r\dot{R}}{A_0}\right)^2\right]\right)_{\Sigma},\label{eq25}\\
\left[p_r\right]_{\Sigma} &=& \left[q f(t)B_0\right]_{\Sigma},\label{eq26}\\
m(r,t) &\stackrel{\Sigma}{=}& m(v). \label{eq27}
\end{eqnarray}
From Eq.~(\ref{eq25}) and (\ref{eq27}), the mass enclosed within the boundary surface at any instant $t$ within a boundary surface $r\leq r_\Sigma$ can be written as
\begin{equation}
m(r,t) = \frac{r f(t)}{2}\left[\frac{r^2}{R^2+r^2}+\left(\frac{r\dot{f}}{A_0}\right)^2\right].\label{eq28}
\end{equation}
Combining Eqs.~(\ref{eq6}), (\ref{eq9}) and (\ref{eq26}), together within the condition $(p_r)_s(r=r_{\Sigma}) = 0$, we deduce the surface equation governing the collapsing matter in the form
\begin{equation}
2\ddot{f}f+\dot{f}^2-2n\dot{f} = 0,\label{eq29}
\end{equation}
where, we have defined
\begin{equation}
n = \left[\frac{A_0'}{B_0}\right]_\Sigma.\label{30}
\end{equation}
Note that for a given initial static configuration $n$ appears as a constant in Eq.~(\ref{eq29}). Following Bonnor\cite{Bonnor}, we write Eq.~(\ref{eq29}) as a first order differential equation 
\begin{equation}
\dot{f} = -\frac{2n}{\sqrt{f}}(1-\sqrt{f}),\label{eq31}
\end{equation}
which admits a solution
\begin{equation}
t = \frac{1}{n}\left[\frac{f}{2}+\sqrt{f}+\ln(1-\sqrt{f})\right].\label{eq32}
\end{equation}
Note that at $t \rightarrow - \infty$, i.e., at the onset of collapse, $f = 1$ and $f(t) \rightarrow 0$ as $t \rightarrow 0$. 

We, therefore, have a complete description of the interior and exterior space-times of the collapsing body. In the following section, we  shall analyze the impact of anisotropy by making use of the solutions thus obtained.    

\section{\label{sec5} Physical analysis:}
To understand the nature of collapse, we assume that a star starts its collapse at $t=-\infty$ (i.e., $f(t)=1$) and the initial static star is described by the parameters $A_0$, $B_0$ and $\Delta_s$. We assume that the collapsing object has an initial radius $r_{\Sigma}(t\rightarrow -\infty) = r_s$ and mass $m(r_s,-\infty) = m_s$ satisfying the condition $2m_s/r_s < 1$. If for some reason, instability develops inside the star, it begins to collapse.  The comoving boundary surface $(rf)_{\Sigma} = r_s f(t)$ then starts shrinking until it reaches its Schwarzschild horizon value $\left[rf(t_{bh})\right]_{\Sigma} = r_s f(t_{bh}) = 2m(v)$, where $t=t_{bh}$ denotes the time of formation of the black hole corresponding to the value of $f(t)=f_{bh}$.  

From Eq.~(\ref{eq28}), the mass of the evolving star at any instant $t$ within the boundary radius $r_{\Sigma}$ may be written as
\begin{equation}
m(r_{\Sigma},t) = \left[m_s f + \frac{2n^2r^3}{A_0 ^2}(1-\sqrt{f})^2\right]_{\Sigma},\label{eq33}
\end{equation}
where, mass of the initial static star has the form 
\begin{equation}
m_s = \int_0^{r_s} 4\pi r^2 \rho_s dr =  \frac{r_s^3}{2(r_s^2 +R^2)}. \label{eq34} 
\end{equation} 
Consequently, the condition $(r f_{bh})_{\Sigma} = 2m(v)$ yields
\begin{equation}
f_{bh} = \left[\frac{\frac{2n r}{A_0}}{\frac{2n r}{A_0} +\sqrt{1-\frac{2m_s}{r}}}\right]^2_{\Sigma},\label{eq35}
\end{equation}
and the time of black hole formation is obtained as 
\begin{equation}
t_{bh} = \frac{1}{n}\left[\frac{f_{bh}}{2}+\sqrt{f_{bh}} +\ln(1-\sqrt{f_{bh}})\right].\label{eq36}
\end{equation} 
The model parameters (namely, $R$, $P$, $Q$) involving the initial static star are determined using the following boundary conditions:
\begin{eqnarray}
A_0(r_s) &=& \left(1-\frac{2m_s}{r_s}\right)^{1/2},\label{eq37}\\
B_0(r_s) &=& \left(1-\frac{2m_s}{r_s}\right)^{-1/2}, \label{eq38}\\
(p_r)_s (r_s) &=& 0,\label{eq39}
\end{eqnarray}
where we have matched the static interior space-time to the Schwarzschild exterior and imposed the condition that the radial pressure ($(p_r)_s$) must vanish at the boundary.

\begin{figure}
\includegraphics[width=0.75\textwidth]{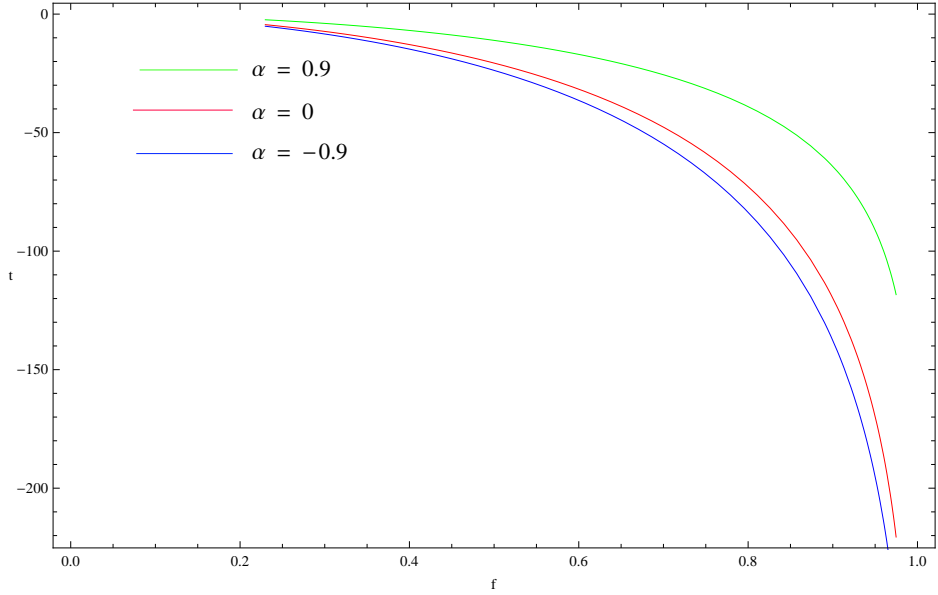}
\caption{Variation of $f(t)$ for different anisotropic parameters.}
\label{fig1}
\end{figure}

Now, to get an insight about the effects of anisotropy, we  consider different initial static configurations, both isotropic $\alpha =0$ and anisotropic ($\alpha \neq 0$). We consider different initially static stellar configurations of identical initial masses and radii (we have assumed $m_s = 3.25~M_{\odot}$ and radius $r_s = 20~$km) with different values of the anisotropic parameter $\alpha$. Using numerical procedures, we have calculated the corresponding model parameters and also evaluated the time of formation of black holes $t_{bh}$ and radius ($r_{bh} =r_s f_{bh}$) and mass ($m_{bh}$) of the black hole formed. Our results have been compiled in Table \ref{tab:table1}. Variations of $f(t)$ for different choices of the anisotropic parameter $\alpha$ have been shown in Fig.~\ref{fig1}. We have also calculated the rate of collapse in our model for different anisotropic parameters. The collapse rate, in our model, is obtained as
\begin{equation}
\Theta  = u_{;\beta}^{\beta} = \frac{3\dot{f}}{A_0 f} = \frac{6n(\sqrt{f}-1)}{A_0 f\sqrt{f}}.\label{eq40}
\end{equation}
The collapse rate turns out to be $\Theta = - 0.101464,~ - 0.102096,~ - 0.100749$ for $\alpha = 0,~0.9,~-0.9$, respectively.  Since, the collapsing object contracts in size as time evolves $\Theta$, in our construction, turns out to be negative. However, from the absolute values of $\Theta$, we note that for a positive anisotropy ($p_r > p_t$) the collapse rate increases as compared to an isotropic star while for a negative value of $\alpha$ ($p_t > p_r$), the rate decreases.
\begin{table}
\caption{\label{tab:table1} Data for collapsing systems having different values of the anisotropic parameter $\alpha$. All the configurations start collapsing with initial masses $m_s = 3.25~M_{\odot}$ and radii $r_s = 20~km$.}
\begin{ruledtabular}
\begin{tabular}{ccccccccc}\\
$\alpha$ & $P$ & $Q$ & $R ~(km)$ & $n$ & $f_{bh}$ & $t_{bh}$ &  $r_{bh}~(km)$ & $m_{bh}~(M_{\odot})$\\ \hline
0.9 & 0.2848 & -0.6341 & 20.8427 & 0.0244 & 0.2298 & -2.3914 & 4.596 & 2.2980 \\
0.6 & 0.3800 & -0.4994 & 20.8427 & 0.0163 & 0.2298 & -3.5816 & 4.596 & 2.2980 \\
0.3 & 0.4065 & -0.4619 & 20.8427 & 0.0142 & 0.2298 & -4.1074 & 4.596 & 2.2980 \\
0 & 0.42067 & -0.4418 & 20.8427 & 0.0131 & 0.2298 & -4.4578 & 4.596 & 2.2980 \\
-0.3 & 0.4300 & -0.4287 & 20.8427 & 0.0124 & 0.2298 & -4.7239 & 4.596 & 2.2980 \\
-0.6 & 0.4367 & -0.4192 & 20.8427 & 0.0118 & 0.2298 & -4.9401 & 4.596 & 2.2980 \\
-0.9 & 0.4418 & -0.4120 & 20.8427 & 0.0114 & 0.2298 & -5.1230 & 4.596 & 2.2980 \\
-1.5 & 0.4493 & -0.4012 & 20.8427 & 0.0108 & 0.2298 & -5.4231 & 4.596 & 2.2980 
\end{tabular}
\end{ruledtabular}
\end{table}
From Table~\ref{tab:table1}, we note that
\begin{itemize}
\item Case I: when $\alpha > 0$ (implying $p_r > p_t$), the horizon is formed at a faster rate as compared to $\alpha = 0$, i.e., isotropic case.
\item Case II: When $\alpha < 0$ (implying $p_t > p_r$), the horizon is formed at a slower rate as compared to $\alpha = 0$, i.e., isotropic case.
\end{itemize}  
Similar observations may be found in Ref.~\cite{Herrera3}. However, we note that mass and radius of the collapsed configuration do not depend on anisotropy in our formulation.

\section{\label{sec6} Thermal behaviour}
\begin{figure}
\includegraphics[width=0.75\textwidth]{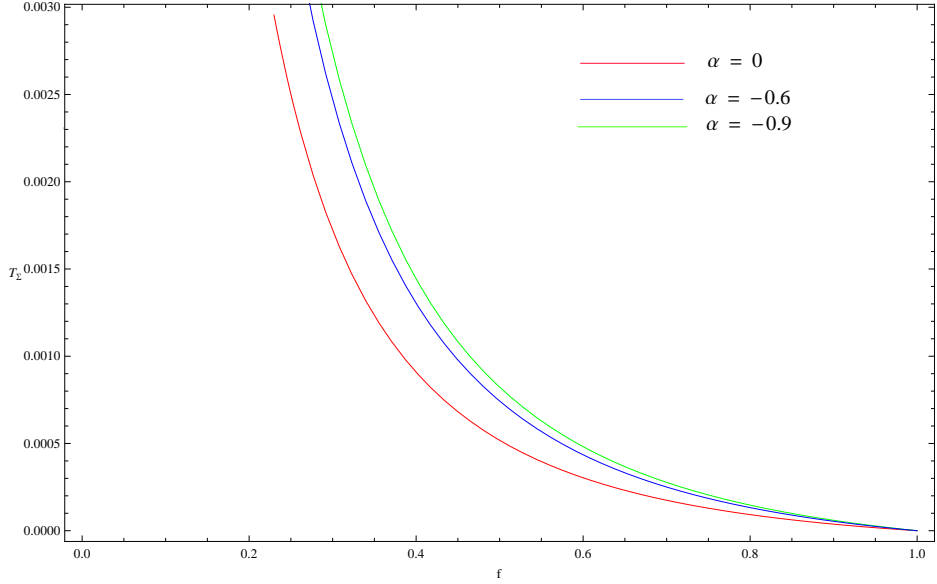}
\caption{Evolution of surface temperature for different anisotropic parameters.}
\label{fig2}
\end{figure}

To analyze the impact of anisotropy on the evolution of temperature of the collapsing system, we use the relativistic Maxwell-Cattaneo relation for temperature governing the heat transport\cite{Israel,Maartens,Martinez} given by
\begin{equation}
\tau(g^{\alpha\beta}+u^{\alpha}u^{\beta})u^{\delta}q_{\beta;\delta} + q^{\alpha} = -\kappa(g^{\alpha\beta}+u^{\alpha}u^{\beta})[T_{,\beta}+T\dot{u_{\beta}}],\label{eq41}
\end{equation}
where $\kappa (\geq 0)$ is the thermal conductivity and $\tau (\geq 0)$ is the relaxation time. For the line element (\ref{eq1}), Eq.~(\ref{eq41}) reduces to
\begin{equation}
\tau\frac{d}{dt}(qfB_0) + q^1 f A_0 B_0 = -\kappa \frac{1}{f B_0}\frac{d}{dr}(A_0T)\label{eq42}
\end{equation}
Following an earlier work\cite{Sharma1}, we write the relativistic Fourier heat transport equation by setting $\tau = 0$ in (\ref{eq42}). For $\tau=0$, combining Eqs.~(\ref{eq6}) and (\ref{eq42}), we get
\begin{equation}
8\pi\kappa(A_0T)' = \frac{2A_0'\dot{f}}{A_0f},\label{eq43}
\end{equation}
Let us now assume that the thermal conductivity varies as $\kappa = \gamma T^{\omega}$, where $\gamma$ and $\omega$ are constants. Eq.~(\ref{eq43}), then  yields
\begin{equation}
8\pi\gamma(A_0T)'=\frac{2A_0' T^{-\omega}}{A_0}\left[\frac{2n(\sqrt{f}-1)}{f\sqrt{f}}\right],\label{eq44}
\end{equation} 
where we have used Eq.~(\ref{eq31}). Integrating the above equation, we get
\begin{equation}
T^{\omega+1}=\frac{n(\sqrt{f}-1)}{2\pi\gamma f\sqrt{f}}\left(\frac{\ln A_0}{A_0}\right)+T_0(t)
,\label{eq45}
\end{equation} 
To get a simple estimate of temperature evolution, we set $\omega=0$, $\gamma = 1$ and $T_0(t) = 0$ and evaluate the surface temperature at any instant as
\begin{equation}
T(r_\Sigma,t)=\frac{n(\sqrt{f}-1)}{2\pi f\sqrt{f}}\left(\frac{\ln A_0}{A_0}\right)_\Sigma
,\label{eq46}
\end{equation}
Time evolution of the surface temperature for different anisotropic parameter $\alpha$ has been shown in Fig.~(\ref{fig2}), where we have used the data from Table~\ref{tab:table1}.

\section{\label{sec7} Discussions}
We have developed a minimalistic and basic framework of a gravitationally collapsing system which has allowed us to examine the impact of pressure anisotropy explicitly. In our construction, we have ignored the effects of various other factors relevant for collapsing systems such as shear, viscosity, charge etc. Moreover, the line element describing the interior space-time has been assumed to be spherically symmetric where the metric functions have been chosen to be separable in its variables. Though formulation of a more general framework is always preferred to describe a realistic situation, the simple model developed here provides a mechanism to capture the impact of anisotropy on gravitational collapse successfully. Though, as pointed out in \cite{Iva2}, effects of factors like shear, charge etc., on self-bound systems can be absorbed by considering the system to be anisotropic, in general, we intend to formulate a more general framework so as to examine the combined impacts of various factor relevant for gravitationally collapsing systems. These issues will be taken up elsewhere.

\begin{acknowledgments}
RS is thankful to the Inter University Centre for Astronomy and Astrophysics (IUCAA), Pune, India, where a part of this work was carried out under its Visiting Research Associateship Programme.
\end{acknowledgments}

\end{document}